\documentclass[10pt,conference]{IEEEtran}
\usepackage{amsmath,amsfonts}
\usepackage{array}
\usepackage[caption=false,font=normalsize,labelfont=sf,textfont=sf]{subfig}
\usepackage{textcomp}
\usepackage{stfloats}
\usepackage{url}
\usepackage{verbatim}
\usepackage{graphicx}
\usepackage{cite}
\usepackage{acronym}

\usepackage{ifpdf}
\usepackage{url}
\ifpdf
\else
\fi

\usepackage{cite}
\usepackage{balance}
\usepackage{bm,comment,color}

\usepackage{amsmath}

\usepackage{algorithm}
\usepackage{algpseudocode}% http://ctan.org/pkg/algorithmicx
\algtext*{EndWhile}% Remove "end while" text
\algtext*{EndIf}% Remove "end if" text
\algtext*{EndFor}

\usepackage{array}
\usepackage{amsfonts} 
\usepackage{amssymb}
\usepackage{esint} % various fancy integral symbols
\usepackage{units}
\usepackage{multirow}
\usepackage{comment}

\usepackage{color}
\usepackage{pgfplots}
\usepackage{tikz}
\usetikzlibrary{calc}
\makeatletter
\newcommand{\gettikzxy}[3]{%
  \tikz@scan@one@point\pgfutil@firstofone#1\relax
  \edef#2{\the\pgf@x}%
  \edef#3{\the\pgf@y}%
}
\usetikzlibrary{spy,backgrounds}
\pgfplotsset{compat=newest}
\usetikzlibrary{plotmarks}
\usetikzlibrary{arrows.meta}
\usepgfplotslibrary{patchplots}
\usepackage{grffile}
\newlength\fheight 
\newlength\fwidth 
\usepgfplotslibrary{fillbetween}

\acrodef{3gpp}[3GPP]{3rd Gneration Partnership Project}
\acrodef{ad}[AD]{autonomous drive}
\acrodef{adas}[ADAS]{advanced driver assistance system}
\acrodef{aoa}[AOA]{angles-of-arrival}
\acrodef{aod}[AOD]{angles-of-departure}
\acrodef{aosa}[AOSA]{array-of-subarray}
\acrodef{bs}[BS]{base station}
\acrodef{bse}[BSE]{beam squint effect}
\acrodef{cdf}[CDF]{cumulative distribution function}

\acrodef{coa}[COA]{curvature of arrival}
\acrodef{crb}[CRB]{Cram\'er-Rao bound}

\acrodef{dbscan}[DBSCAN]{density-based spatial clustering of applications with noise}
\acrodef{elaa}[ELAA]{extremely-large antenna array}
\acrodef{ff}[FF]{far field}
\acrodef{fim}[FIM]{Fisher information matrix}
\acrodef{gnss}[GNSS]{global navigation satellite system}
\acrodef{gps}[GPS]{global positioning system}
\acrodef{imu}[IMU]{inertial measurement unit}
\acrodef{ip}[IP]{incidence point}
\acrodef{kld}[KLD]{Kullback–Leibler divergence}
\acrodef{las}[L\&S]{localization and sensing}
\acrodef{los}[LOS]{line-of-sight}
\acrodef{mae}[MAE]{mean absolute value}
\acrodef{map}[MAP]{maximum a posteriori}
\acrodef{mle}[MLE]{maximum likelihood estimator}
\acrodef{mm}[MM]{mismatched model}
\acrodef{mpc}[MPC]{multipath component}
\acrodef{nlos}[NLOS]{non-line-of-sight}
\acrodef{nf}[NF]{near field}
\acrodef{nr}[NR]{new radio}
\acrodef{ofdm}[OFDM]{orthogonal frequency division multiplexing}
\acrodef{pbd}[PBD]{partial blockage detection}
\acrodef{prs}[PRS]{positioning reference signal}
\acrodef{psd}[PSD]{power spectral density}
\acrodef{pss}[PSS]{primary synchronization signal}
\acrodef{rmse}[RMSE]{root mean squared error}
\acrodef{rf}[RF]{radio frequency}
\acrodef{rfc}[RFC]{radio frequency chain}

\acrodef{ris}[RIS]{reconfigurable intelligent surface}
\acrodef{rss}[RSS]{received signal strength}
\acrodef{rtk}[RTK]{real-time kinematic}
\acrodef{rtt}[RTT]{round-trip-time}
\acrodef{sa}[SA]{sub-array}
\acrodef{simo}[SIMO]{single-input-multiple-output}
\acrodef{slam}[SLAM]{simultaneous localization and mapping}
\acrodef{sns}[SNS]{spatial non-stationarity}
\acrodef{sp}[SP]{scattering point}
\acrodef{ssb}[SSB]{synchronization signal/physical broadcast channel block}
\acrodef{swm}[SWM]{spherical wave model}
\acrodef{tdd}[TDD]{time division duplex}
\acrodef{tdoa}[TDOA]{time-difference-of-arrival}
\acrodef{tm}[TM]{true model}
\acrodef{toa}[TOA]{time-of-arrival}
\acrodef{ue}[UE]{user equipment}
\acrodef{ura}[URA]{uniform rectangular array}
\acrodef{va}[VA]{virtual anchor}

\long\def\comment#1{}

\DeclareMathOperator*{\argmin}{arg\,min}

\newfont{\bbb}{msbm10 scaled 700}

\newfont{\bb}{msbm10 scaled 1100}

\newcommand{\av}{{\bf a}}
\newcommand{\bv}{{\bf b}}
\newcommand{\cv}{{\bf c}}
\newcommand{\dv}{{\bf d}}

\newcommand{\gv}{{\bf g}}
\newcommand{\hv}{{\bf h}}

\newcommand{\mv}{{\bf m}}
\newcommand{\nv}{{\bf n}}

\newcommand{\pv}{{\bf p}}

\newcommand{\sv}{{\bf s}}
\newcommand{\tv}{{\bf t}}
\newcommand{\uv}{{\bf u}}
\newcommand{\wv}{{\bf w}}
\newcommand{\vv}{{\bf v}}

\newcommand{\yv}{{\bf y}}

\newcommand{\Um}{{\bf U}}
\newcommand{\Wm}{{\bf W}}
\newcommand{\Vm}{{\bf V}}

\newcommand{\Ym}{{\bf Y}}

\newcommand{\alphav}{\hbox{\boldmath$\alpha$}}

\newcommand{\thetav}{\hbox{$\boldsymbol\theta$}}
\newcommand{\tauv}{\hbox{\boldmath$\tau$}}

\newcommand{\herm}{{\sf H}}

\begin{document}

\bstctlcite{IEEEexample:BSTcontrol}

\title{ELAA Near-Field Localization and Sensing with Partial Blockage Detection}

\author{
Hui Chen\IEEEauthorrefmark{1},
Pinjun Zheng\IEEEauthorrefmark{2}, 
Yu Ge\IEEEauthorrefmark{1},
Ahmed Elzanaty\IEEEauthorrefmark{3},
Jiguang He\IEEEauthorrefmark{4},
Tareq~Y.~Al-Naffouri\IEEEauthorrefmark{2},
Henk Wymeersch\IEEEauthorrefmark{1}
\\
\IEEEauthorrefmark{1}Chalmers University of Technology, Sweden

\IEEEauthorrefmark{2}King Abdullah University of Science and Technology, KSA
% (\{pinjun.zheng, tareq.alnaffouri\}@kaust.edu.sa)
\\ 
\IEEEauthorrefmark{3}University of Surrey, UK 
% (a.elzanaty@surrey.ac.uk)

\IEEEauthorrefmark{4}Technology Innovation Institute, UAE 
% (jiguang.he@tii.ae)
% (\{hui.chen, yuge, henkw\}@chalmers.se)
% \\
% E-mail: hui.chen@chalmer.se

% \IEEEauthorrefmark{1}Chalmers University of Technology, Sweden\ \ 
% \\
% \IEEEauthorrefmark{2}King Abdullah University of Science and Technology, KSA
% \\
% \IEEEauthorrefmark{3}University of Surrey, UK
% \\
% E-mail: hui.chen@chalmer.se

}
% This work has been supported by the SNS JU project 6G-DISAC under the EU’s Horizon Europe research and innovation programme under Grant Agreement No 101139130.



\maketitle

\begin{abstract}
High-frequency communication systems bring extremely large aperture arrays (ELAA) and large bandwidths, integrating localization and (bi-static) sensing functions without extra infrastructure. 
Such systems are likely to operate in the near-field (NF), where the performance of localization and sensing is degraded if a simplified far-field channel model is considered. 
However, when taking advantage of the additional geometry information in the NF, e.g., the encapsulated information in the wavefront, localization and sensing performance can be improved. In this work, we formulate a joint synchronization, localization, and sensing problem in the NF. Considering the array size could be much larger than an obstacle, the effect of partial blockage (i.e., a portion of antennas are blocked) is investigated, and a blockage detection algorithm is proposed. The simulation results show that blockage greatly impacts performance for certain positions, and the proposed blockage detection algorithm can mitigate this impact by identifying the blocked antennas.
\end{abstract}

\begin{IEEEkeywords}
Near field, localization, sensing, partial blockage, extremely large aperture array.
\end{IEEEkeywords}

\section{Introduction}
Localization is expected to be integrated into the communication systems due to the increased bandwidth and array size with mmWave/THz band signals~\cite{behravan2022positioning}. A by-product of localization is the estimated position of surrounding objects or \acp{sp} from \ac{nlos} paths, which is usually called (bi-/multi-static) sensing. In the \ac{3gpp} project, positioning in the 5G \ac{nr} has been studied in TR38.855~\cite{tr38855}, followed by an expanded study in TR38.895~\cite{tr38895}. In the meantime, many theoretical analysis and experimental studies have also been carried out, both showing huge potentials of \ac{las} in communication systems~\cite{gao2022toward, ruan2022ipos, ge2023experimental}. Consequently, such integration can provide vast opportunities for Internet-of-Things applications, augmented reality, smart cities, and other new use cases~\cite{liu2022integrated}.

Due to the preferable simplicity, the mmWave channels are usually modeled with the \ac{ff} assumption, based on which multi-dimensional signal processing algorithms such as ESPRIT can be adopted to extract the channel parameters of interest~\cite{roemer2014analytical,zheng2023jrcup}. The angle/delay information (from phase changes across antennas/subcarriers) of the \ac{los} path and NLOS paths provide the position information of the target and the incident points (e.g., scatter points or objects), respectively \cite{roemer2014analytical}. However, the \ac{ff} assumption is no longer valid for \ac{elaa} and wideband systems \cite{chen20236g, Elzanaty:23}. Communications, as well as \ac{las} tasks that are performed in the \ac{nf} experience channel features such as \ac{sns}, \ac{bse}, and \ac{swm}~\cite{chen20236g}. 

When adopting a \ac{ff} model that ignores these features, model mismatch happens, and \ac{las} performance will be affected~\cite{chen2022channel}. However, when considering these features, extra geometrical information can be obtained, although at the cost of increased channel model complexity.
For example, localization in the \ac{nf} can be performed with narrowband signals by exploiting the \ac{coa}~\cite{guerra2021near}, even under LOS blockage~\cite{ozturk2023ris,Rinchi:22}. When considering wideband signals, joint localization and synchronization can be performed~\cite{dardari2021nlos}. The \ac{bse} can also be utilized for coverage enhancement thanks to the spatial diversity spanned by different subcarriers (i.e., squinted beams)~\cite{cui2022near}. Hence, involving more accurate \ac{nf} channel models and developing corresponding signal processing algorithms are needed. 

\begin{figure}[t]
% \begin{minipage}[b]{0.78\linewidth}
\centering
\centerline{\includegraphics[width=0.7\linewidth]{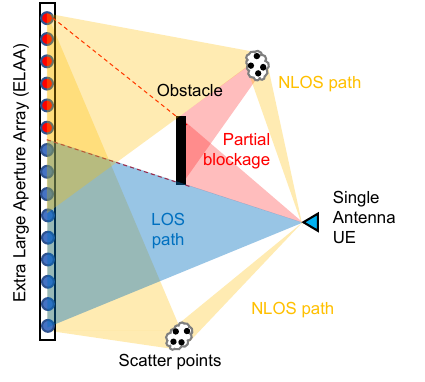}}
\caption{Illustration of localization and sensing under partial blockage.}
% (\textit{Figure elements by macrovector on Freepik})} 
\vspace{-5mm}
\label{fig_illustration}
\end{figure}

In addition to the \ac{nf} model, another issue accompanying the large array is the partial blockage of the antennas. Unlike the \ac{ff} model where the antenna array is treated as a point target, and the whole antenna's blockage is considered if an object exists between the transmitter and the receiver, partial blockage indicates that a portion of antennas is affected. Similar research works have studied antenna diagnosis~\cite{sun2021blind} and RIS pixel failures~\cite{ozturk2023ris}, where only few antennas are impaired. However, a larger portion of antennas might be affected by blockage, depending on the position and shape of the obstacles between the transmitter and receivers, making the detection problem challenging for analog arrays. 
% When ignoring these blockages, model mismatch happens. However, if the blocked antennas can be detected, localization performance can be improved compared to the case without the blockage, and passive sensing of the obstacle becomes also possible.

In this work, we study the \ac{las} in the \ac{nf} and consider partial blockage of the \ac{elaa}. The contributions of this work can be summarized as follows: (i) a \ac{nf} joint synchronization, localization, and sensing problem under a stationary scenario is formulated in an uplink \ac{simo} system with a single \ac{bs}, which was previously not possible using a \ac{ff} model; (ii) the effect of partial blockage on localization is evaluated based on the \ac{kld} metric; (iii) a \ac{pbd} algorithm is developed to detect blocked antennas in an analog or hybrid array with a limited number of transmissions.

% Contribution:

% 1. We formulate the problem of near-field localization under partial blockage and evaluate the performance loss when such a blockage is ignored.

% 2. We proposed a partial blockage detection algorithm for analog arrays.

% 3. We proposed a multi-stage algorithm for joint localization, sensing, and synchronization.

% 4. Simulations are carried out to evaluate the derived localization bound and blockage detection algorithm.

% Test of references~\cite{elzanaty2021reconfigurable}. 
% \newpage
\section{System Model}
% \subsection{Geometry Model}
This section describes the signal and channel models of the localization system. We consider a 2D uplink \ac{simo} system with $L$ \acp{sp}. An $N$-antenna \ac{bs} is located along the $y$-axis with its center denoted by $\pv_{\text{B}}$, and each antenna is located at $\bv_n$ ($n = 1, \ldots, N$). The \ac{ue} and the $\ell$-th \ac{sp} are located at $\pv_{0}$ and $\pv_{\ell}$, respectively. The obstacle may appear between the transmitter and the receiver and block part of the array in the \ac{los} and \ac{nlos} paths, as seen in Fig.~\ref{fig_illustration}. For the convenience of performance analysis, we start with a simple scenario by adopting a point-target model of the \ac{sp}. More realistic and complicated models, such as circular targets or extended targets~\cite{garcia2022cramer} can be considered in future works.

\subsection{Signal Model}
An~\ac{ofdm}-based transmission is adopted in this system with $K$ subcarriers and $G$ transmissions. Given the transmitted signal symbol for the $k$-th subcarrier and $g$-th transmission $x_{g,k}$ ($|x_{g,k}| = \sqrt{P}$ with $P$ as the average transmit power), the received signals can be expressed as 
\begin{equation}
    \yv_{g,k} = \Wm_g \hv_k x_{g,k} + \Wm_g \nv_{g,k}.
    \label{eq_signal_model}
\end{equation}
Here, $\yv_{g,k} \in \mathbb{C}^{M}$ with $M$ as the number of \acp{rfc}, $\hv_k \in \mathbb{C}^{N}$ is the channel vector of the $k$-th subcarrier that is assumed to be coherent during the $G$ transmissions, $\nv_{g,k} \in \mathcal{CN}(\mathbf{0}_N, N_0 W \mathbf{I}_N)$ is the noise vector with $N_0$ as the noise \ac{psd} and $W$ as the bandwidth, $\Wm_g \in \mathbb{C}^{M\times N}$ is the combining matrix that differs depending on array architectures.\footnote{For digital arrays, $\Wm_g = \mathbf{I}_N$ ($M=N$) and can thus be removed. For fully connected hybrid arrays consisting $M_\text{R}$ \acp{rfc}, $\Wm_g \in \mathbb{C}^{S\times N}$ with each element $|w_{i,j}| = \frac{1}{\sqrt{N}}$.}
For \ac{aosa} structures and assume $N_\text{S}$ is the number of antennas for all the $S$ \acp{sa} ($N = S N_\text{S}$), the combiner can be expressed as $\Wm_g = \text{blkdiag}(\wv_{g,1}^\top, \ldots, \wv_{g,S}^\top) \in \mathbb{C}^{S\times N}$ with $\wv_{g,s} \in \mathbb{C}^{S}$ ($s = 1, \ldots, S$) as the combiner vector of each \ac{sa}, $(\cdot)^\top$ as the transpose operation, and $\text{blkdiag}(\cdot)$ indicating the block diagonal operation. In this work, we adopt the \ac{aosa} structure with $S$ \acp{sa} with each of them connected to a specific \ac{rfc} (the number of \acp{rfc} $M$ equals to $S$).

\subsection{Far Field Channel Model}
When considering the \ac{ff} scenario, the channel vector for the $k$-th subcarrier $\hv_k^{\text{FF}} \in \mathbb{C}^{{N}}$ can be formulated as
\begin{align}
    \hv_{k}^{\text{FF}} = & \sum_{\ell = 0}^{L}\hv_{\ell,k}^{\text{FF}} = \sum_{\ell = 0}^{L}\alpha_{\ell}\, \av_\ell(\vartheta_\ell) D_k(\tau_\ell),
    \label{eq:far_field_channel_model}
\end{align}
where $L$ is the number of paths, $\ell=0$ indicates the LOS path and $\ell>0$ corresponds to the $\ell$-th NLOS path. For the $\ell$-th path ($\ell = 0, \ldots, L$), $\alpha_{\ell}$, $\av_{\ell}(\vartheta_{\ell})$, $D_k(\tau_{\ell})$ are the complex channel gain, \ac{ff} steering vector, and delay component at the $k$-th subcarrier, respectively, with $\vartheta_{\ell}$ and $\tau_\ell$ as the \ac{aoa} and delay of the $\ell$-th path. More specifically,

\begin{align}
    \alpha_{\ell} = & \rho_{\ell} e^{-j\xi_{\ell}} 
    = 
    \begin{cases}
    \frac{\lambda_c e^{-j\xi_{\ell}}}{4\pi \Vert \pv_{_0} - \pv_{_\text{B}} \Vert} & \ell=0, 
    \\
    \sqrt{\frac{c_{\ell}}{4\pi}}\frac{\lambda_c e^{-j\xi_{\ell}}}{4\pi \Vert \pv_{_0} - \pv_{\ell} \Vert \Vert \pv_{\ell} - \pv_{_\text{B}} \Vert} & \ell > 0, 
    \end{cases}
    \label{eq:far_field_gain}
    \\
    \av_\ell(\vartheta_{\ell}) = &[e^{-j\pi\frac{N-1}{2}\sin(\vartheta_{\ell})}, \ldots, 1, \ldots, e^{-j\pi\frac{1-N}{2}\sin(\vartheta_{\ell})}]^\top,
    \label{eq:far_field_steering_vector}
    \\
    D_k(\tau_{\ell}) = & e^{-j \frac{2 \pi}{\lambda_k} \tau_{\ell}} = 
    \begin{cases}
    e^{-j\frac{2\pi}{\lambda_k}(\Vert \pv_{_0} - \pv_{_\text{B}} \Vert+\beta)} & \ell=0,
    \\
    e^{-j\frac{2\pi}{\lambda_k}(
    \Vert \pv_{_0} - \pv_{\ell} \Vert + 
    \Vert \pv_{\ell} - \pv_{_\text{B}} \Vert +
    \beta)} & \ell>1,
    \end{cases}
    % D_k(\pv, \beta) = e^{-j\frac{2\pi}{\lambda_k}(\Vert \pv_{_\text{U}} - \pv_{_\text{B}} \Vert+\beta)},
    \label{eq:far_field_delay_component}
\end{align}
where $\lambda_c$ and $\lambda_k$ are the wavelengths corresponding to the central frequency and the $k$-th subcarrier, and $c_\ell$ is the Radar cross section of the $\ell$-th path. Due to the unknown clock offset $\beta$ in $\tau_{\ell}$, localization cannot be performed in a \ac{simo} system with a single \ac{bs} under \ac{ff} scenarios, even with wideband signals.

\subsection{Near-field Channel Model}
By adopting a \ac{nf} model, the channel vector $\hv_{k}^{\text{NF}}$ can then be formulated as
% Some test: the estimator seems to be biased (probably because of the initial point and the GD algorithm?), does it matter?
% The difference between \ac{nf} and FF channel models is in the channel vector in~\eqref{eq:far_field_channel_model}. The phase change caused by delay can no longer be expressed using a steering vector and the amplitudes for different frequencies at different antennas are varying. 
\begin{align}
\begin{split}
% \hv_{k}^{\text{NF}} = & \sum_{\ell=0}^L \hv_{\ell, k}^{\text{NF}}\\
    \hv_{k}^{\text{NF}} = \sum_{\ell=0}^L \alphav_{\ell,k}(\pv_\ell) \odot \mv_{\ell} \odot \dv_{\ell,k}(\pv_\ell) D_{\ell, k}(\pv_0, \pv_\ell, \beta),
    \label{eq:near_field_channel_model} 
\end{split}
\end{align}
where $\mv_{\ell}\in\mathbb{R}^N$ is the masking vector with each element as either 1 (unblocked) or 0 (blocked). {Diffraction may happen depending on the shape and materials of the object and cause non-zero coefficients for the blocked antennas~\cite{han2014multi}, which will be evaluated in the simulation.} The frequency-dependent \ac{nf} channel gain, steering vector, and delay component $\alphav_{\ell,k}$, $\dv_{\ell,k}$, and $D_{\ell,k}$ can be expressed as 
\begin{align}
    \alpha_{\ell,k,n}(\pv_\ell) = & \alpha_{\ell}  c_{\ell, k,n}(\pv_\ell), 
    \ \ 
    c_{\ell, k,n}(\pv_\ell)= 
    \frac{\lambda_k \Vert \pv_{\ell} - \pv_{_\text{B}} \Vert}{\lambda_c \Vert \pv_{\ell}-\bv_n \Vert},
    \label{eq:near_field_channel_amplitude}
    \\
    d_{\ell, k,n}(\pv_\ell) = & e^{-j \frac{2\pi}{\lambda_k} (\Vert \pv_{\ell} -\bv_n \Vert - \Vert \pv_{\ell} - \pv_{_\text{B}} \Vert)},
    \label{eq:near_field_steering_vector}
    \\
    D_{\ell, k}(\pv_0, \pv_\ell, \beta) = &
    \begin{cases}
    e^{-j\frac{2\pi}{\lambda_k}(\Vert \pv_{_0} - \pv_{_\text{B}} \Vert+\beta)} & \ell=0,
    \\
    e^{-j\frac{2\pi}{\lambda_k}(
    \Vert \pv_{_0} - \pv_{\ell} \Vert + 
    \Vert \pv_{\ell} - \pv_{_\text{B}} \Vert +
    \beta)} & \ell>1.
    \end{cases}
    \label{eq:near_field_delay_component}
\end{align} 
Note that equations \eqref{eq:near_field_delay_component} and \eqref{eq:far_field_delay_component} are identical, and the near-field features are reflected on~\eqref{eq:near_field_channel_amplitude} and \eqref{eq:near_field_steering_vector}.

% \section{Performance Lower Bound}
% \red{May not be needed; refer to the MCRB GLOBECOM works.}

\section{Localization and Sensing}
In this section, we describe the \ac{las} algorithms that ignore the blockage of the antennas (i.e., assuming no blockage with $\mv_\ell=\mathbf{1}_N$), and the \ac{pbd} is detailed in Sec.~\ref{sec_pbd}. The \ac{las} problem can be formulated as estimating the state vector $\sv_\text{N} = [\alpha_0, \pv_0^\top, \ldots, \alpha_L, \pv_L^\top, \beta]^\top$ from the observed signal vector $\yv = [\yv^\top_{1}, \ldots, \yv^\top_{S}]^\top \in \mathbb{C}^{SGK}$ of an \ac{aosa} system with $\yv_{s} = [y_{1,1,s}, \ldots, y_{1,K,s}, y_{2,1,s}, \ldots, y_{G,K,s}]^\top \in \mathbb{C}^{GK}$. Usually, the complex channel gains $\alpha_{\ell}$ ($\ell = 0, \ldots, L$) are treated as nuisance parameters, and the nuisance-free state vector can be expressed as $\sv = [\pv_0^\top, \ldots, \pv_L^\top, \beta]^\top$.
Next, we propose a multi-stage \ac{las} algorithm to provide coarse positions of the \ac{ue} and \acp{sp}, and then perform position refinement using \ac{mle}. 
% \red{no blockage, association issue due to low resolution of subarrays}
% Although the \ac{nf} features can provide additional location information, these features require direct localization 

\subsection{Coarse Localization Algorithm}
\subsubsection{Localization with only LOS Path}
By segmenting the whole large array into $S$ \acp{sa}, we can adopt available FF algorithms to estimate the \ac{aoa} $\theta_s$ and delay $\tau_s$ of the $s$-th \ac{sa} separately~\cite{wymeersch2020fisher}.
For the $s$-th \ac{sa}, the estimation problem can be formulated as~\cite{chen2023modeling}
\begin{equation}
    [\hat \theta_{s}, \hat \tau_{s}] = \argmin_{\theta, \tau}
    \left\Vert 
    \yv_s - \frac
    {\uv_s^\herm(\theta, \tau)\yv_s}
    {\Vert \uv_s(\theta, \tau) \Vert ^2}\uv_s(\theta, \tau)
    \right\Vert,
\label{eq_coarse_channel_estimation}
\end{equation}
where $\uv_s(\theta, \tau) = \text{vec}(\Um_s(\theta, \tau)) \in \mathbb{R}^{GK}$ with each entry of $\Um_s(\theta, \tau)$ as $U_{s,g,k} = \wv_g^\top \av(\theta) D_k(\tau)x_{g,k}$ defined in~\eqref{eq:far_field_steering_vector} and~\eqref{eq:far_field_delay_component}, and $(\cdot)^\herm$ is the Hermitian operation. Note that the \ac{aoa} $\theta$ and delay $\tau$ are calculated based on the center of each \ac{sa} denoted as $\pv_{\text{SA}, s}$.

Once the channel parameters for each path are estimated, we can then obtain a coarse location estimate by minimizing the distances between the UE position and all the \ac{aoa} direction vectors, as
\begin{equation}
\hat \pv_{0} = \argmin_{\pv} \sum_{s=1}^{S}\Vert \pv_{\text{SA}, s} +\tv(\hat\theta_s)^\top(\pv- \pv_{\text{SA}, s}) \tv(\hat\theta_s)   - \pv \Vert.
\label{eq_coarse_PU_SA}
\end{equation}
Here, $t(\theta) = [\cos(\theta), \sin(\theta)]^\top$. A coarse clock offset $\hat \beta$ can also be obtained as
\begin{equation}
    \hat \beta = \frac{1}{S}\sum_{s=1}^S (\hat\tau_s - \Vert \pv_{\text{SA}, s} - \hat\pv_0 \Vert).
\label{eq_coarse_clock_offset}
\end{equation}

Finally, based on the coarse estimation from (10) and (11), an \ac{mle} problem for LOS-only NF localization can be formulated as
\begin{equation}
    [\hat \pv_0, \hat \beta] = \argmin_{\pv, \beta}
    \Vert 
    \yv - \frac
    {\vv_0^\herm(\pv, \beta)\yv}
    {\Vert \vv_0(\pv, \beta) \Vert ^2}\vv_0(\pv, \beta)
    \Vert,
\label{eq_coarse_UE_position}
\end{equation}
with $\vv_0(\pv, \beta) = [\text{vec}(\Vm_{0,1}(\pv, \beta))^\top, \ldots, \text{vec}(\Vm_{0, S}(\pv, \beta))^\top]$ as the normalized noise-free received signal in a similar structure as $\yv$, where each entry of $\Vm_{0, s}(\pv, \beta)$ is $V_{0,s,g,k} = \wv_g^\top (\cv_{k}(\pv)\odot \dv_k(\pv)) D_k(\pv, \beta)x_{g,k}$.

\subsubsection{Localization and Sensing with Strong NLOS Paths}
When signals from different paths are resolvable, multi-dimensional algorithms such as ESPRIT can be applied to estimate the channel parameters of each path directly~\cite{wen20205g}. By picking up the shortest path (i.e., minimum delay) as the LOS path, an initial estimate of UE position and clock bias can be obtained based on~\eqref{eq_coarse_PU_SA} and~\eqref{eq_coarse_clock_offset}.
Then, the position of the SPs (assuming correct channel parameters association of each \ac{sp}) can be obtained as
\begin{equation}
    \hat\pv_{\ell} = \argmin_\pv
    \begin{bmatrix}
    \hat \thetav_{\ell} - \thetav_\text{SA}(\pv) \\
    \hat \tauv_{\ell} - \tauv_\text{SA}(\pv) - \hat\beta
    \end{bmatrix}
    \boldsymbol{\mathcal{I}}_{\boldsymbol{\eta}_{\ell}}
    \begin{bmatrix}
    \hat \thetav_{\ell} - \thetav_\text{SA}(\pv) \\
    \hat \tau_{\ell} - \tauv_\text{SA}(\pv) - \hat\beta
    \end{bmatrix}^\top,
    \label{eq_coarse_sp}
\end{equation}
where $\hat\thetav_{\ell} = [\hat\theta_{\ell, 1}, \ldots, \hat\theta_{\ell, S}]^\top$ and $\hat\tauv_{\ell} = [\hat\tau_{\ell, 1}, \ldots, \hat\tau_{\ell, S}]^\top$ are the estimated channel parameters of the $\ell$-th path. Vectors $\thetav_\text{SA}(\pv) = [\theta_{\text{SA},1}(\pv), \ldots, \theta_{\text{SA},S}(\pv)]$ with $\theta_{\text{SA}, s}(\pv) = \text{atan2}(p_1-p_{\text{SA},s, 1}, p_2-p_{\text{SA},s, 1})$, and $\tauv_\text{SA}(\pv) = [\tau_{\text{SA},1}(\pv), \ldots, \tau_{\text{SA},S}(\pv)]$ with $\tau_{\text{SA}, s}(\pv) = \Vert \pv_{\text{SA}, s}(\pv) - \pv - \hat\beta\Vert$ are the angle and delay information at \acp{sa} for a given $\pv$. 
The weighting matrix $\boldsymbol{\mathcal{I}}_{\boldsymbol{\eta}_{\ell}}$ can be selected based on the statistical information of the channel parameters $[\hat \thetav^\top_{\ell}, \hat \tauv^\top_{\ell}]$.
However, the NLOS paths are usually much weaker than the LOS path, and the corresponding \ac{las} algorithm will be presented next.
% Note that this coarse estimation method does not work when the UE is fully or partially blocked such that the LOS path is weaker than the NLOS paths, which is left for future works.

\subsubsection{Localization and Sensing with Weak NLOS Paths}
Considering the NLOS path is much weaker, we need to remove the LOS path from the signal and estimate \acp{sp} from the residual signals defined as
\begin{equation}
    \tilde \yv = \yv - \frac
    {\uv^\herm(\hat\thetav_0, \hat\tauv_0)\yv_s}
    {\Vert \uv(\hat\thetav_0, \hat \tauv_0) \Vert ^2}\uv(\hat\thetav_0, \hat\tauv_0),
    \label{eq_residual_signals}
\end{equation}
where $\hat \thetav_0 = [\hat\theta_{0,1}, \ldots, \hat\theta_{0,S}]$ and $\hat \tau_0 = [\hat\tau_{0,1}, \ldots, \hat\tau_{0,S}]$ contains channel parameters of the LOS path that can be obtained based on~\eqref{eq_coarse_channel_estimation} to \eqref{eq_coarse_UE_position}, $\uv = [\uv_1^\top, \ldots, \uv_S^\top]$ is the concatenation of the normalized noise-free received signal at each \ac{sa}. The estimation of \acp{sp} can be performed based on \eqref{eq_coarse_sp} using the residual signals obtained in~\eqref{eq_residual_signals}.

\subsection{Maximum Likelihood Estimator}
Once a coarse estimation of the state vectors $\sv$ is obtained, we can then formulate a \ac{mle} problem similarly to~\eqref{eq_coarse_UE_position} as
\begin{equation}
\hat \sv = \argmin_\sv 
\Vert 
\yv - \boldsymbol{\Upsilon}(\sv) \gv(\sv)
\Vert,
\label{eq_mle}
\end{equation}
where $\gv(\sv) = (\boldsymbol{\Upsilon}^\herm(\sv) \boldsymbol{\Upsilon}(\sv))^{-1}\boldsymbol{\Upsilon}^\herm(\sv) \yv \in \mathbb{C}^{L+1}$ contains the gain of each identified path, $\boldsymbol{\Upsilon}(\sv)  = [\vv_0(\pv_0, \beta), \ldots, \vv_L(\pv_0, \pv_\ell, \beta)] \in \mathbb{C}^{SGK \times (L+1))}$. 
The $\ell$-th ($\ell >0$) vector in $\boldsymbol{\Upsilon}(\sv)$ can be expressed as $\vv_\ell(\pv_0, \pv_\ell, \beta) = [\text{vec}(\Vm_{\ell, 1}(\pv_0, \pv_\ell, \beta))^\top, \ldots, \text{vec}(\Vm_{\ell,S}(\pv_0, \pv_\ell, \beta))^\top]$, where each entry of the matrix $\Vm_{\ell, s}(\pv_0, \pv_\ell, \beta)$ is given by $V_{\ell, s, g,k} = \wv_g^\top (\cv_{k}(\pv_\ell)\odot \dv_k(\pv_\ell)) D_k(\pv_0, \pv_\ell, \beta)x_{g,k}$.
{The \ac{crb} can be used to benchmark the algorithm performance as detailed in~\cite[Sec. 3.4]{kay1993fundamentals}, which will not be detailed in this work.}

\section{Partial Blockage Detection}
\label{sec_pbd}
Once the position of \ac{ue} and \ac{sp} are estimated, the \ac{pbd} algorithm can be performed. In this section, we focus on the partial blockage of \ac{los} path only and propose a heuristic algorithm to detect the blocked antennas.

\subsection{The Effect of Partial Blockage}
Partial blockage affects localization and sensing performance differently depending on the knowledge of the blockage. If we know the blocked antennas, the \ac{las} performance loss is mainly due to the reduced array size (i.e., only visible antennas provide information). However, if we ignore the blockage and perform
\ac{las} tasks by assuming all the antennas are visible to the UE, model mismatches are introduced, resulting in severe performance degradation.
More specifically, the estimation error can be lower bounded by~\cite{fortunati2017performance}
\begin{equation}
    \text{LB}(\bar \sv, \sv_0) = \text{MCRB}(\sv_0) + (\bar \sv - \sv_0)(\bar \sv - \sv_0)^\top,
    \label{eq_mcrb_bias}
\end{equation}
which contains an MCRB term and a biased term characterized by pseudotrue parameters $\sv_0$, with $\bar\sv$ as the true state~\cite{fortunati2017performance}. The pseudotrue parameter $\sv_0$ is defined as the vector that minimizes the \ac{kld} between the probability density function of the blocked model $f_B$ (signal model from~\eqref{eq_signal_model} and~\eqref{eq:near_field_channel_model}) and the assumed model $f_U$ (i.e., by assuming no blockage with $\mv_\ell = \mathbf{1}_N$) as
\begin{equation}
    \sv_0 = \argmin_\sv D_\text{KL}(f_\text{B}(\yv | \bar \sv) || f_\text{U}(\yv| \sv)).
    \label{eq_pseudotrue_parameter}
\end{equation}
At high SNR, MCRB is ignorable and the biased term dominates, which limits the fundamental mismatched error. In other words, the estimation that ignores the effect of partial blockage will introduce a biased term, resulting in a result $\sv_0$ with mismatch error instead of the true state $\bar \sv$.

\subsection{Partial Blockage Problem Formulation}
\subsubsection{Blockage Detection with Sufficient Observations}
It is easy to detect the partial blockage of the LOS path by simple thresholding in digital arrays. For analog and hybrid arrays, similar approaches can be applied to recover the beamspace channel to the element-space channel with sufficient observations (i.e., no less than the number of antennas in analog arrays).
Take the DFT codebook in an analog array as an example; concatenating the observed signals as $\tilde \yv = [\yv_{1}, \ldots, \yv_{G}]^\top$ and multiplying by an IFFT matrix will result in an element space channel vector (with noise). 
When other combiner vectors are adopted (e.g., $\Wm_g \in \mathcal{R}^{M\times N}$,  $M>1$), this process can be done by multiplying a pseudoinverse matrix of $\tilde \Wm=[\Wm_1, \ldots, \Wm_G]$ to the concatenated observations $\tilde \yv = [\yv_{1,1}, \ldots, \yv_{1,M}, \ldots, \yv_{G,M}]^\top$ to obtain an element-space channel vector, with $\yv_{g,m}$ as the received signal of the $g$-th transmission at the $m$-th \ac{rfc}.

\subsubsection{PBD with a Limited Number of Observations}
When there is only a limited number of observations (i.e., less than the number of antennas in analog arrays), the \ac{pbd} task becomes challenging. If we assume the blocked antennas are adjacent to each other, the detection problem becomes feasible by implementing a heuristic way. Specifically, we want to find the masking vector where a continuous sequence of antennas is all zeros. By considering the blockage of the LOS path (given the fact that the NLOS paths are much weaker), the problem can be formulated as finding the masking vector $\mv$ that minimizes the objective function $J(\hat \sv, \mv)$ given by
\begin{equation}
    \hat \mv = \argmin_{\mv}J(\hat \sv, \mv) = \argmin_{\mv} ||\yv - 
    \tilde{\boldsymbol{\Upsilon}}(\hat \sv, \mv)
    \gv(\hat\sv)||,
    \label{eq_blockage_cost_function}
\end{equation}
where $\tilde{\boldsymbol{\Upsilon}} = [\uv_0(\pv, \beta)\odot \mv, \uv_1, \ldots, \uv_L]$ and $\gv(\hat\sv)$ is defined in~\eqref{eq_mle}. The detection accuracy can be defined as 
\begin{equation}
    \text{Accuracy} = 1 - \frac{\Vert \hat \mv - \bar \mv\Vert^2}{N}.
\end{equation}

\subsection{The Proposed Algorithm}
The detection problem in~\eqref{eq_blockage_cost_function} is an NP-hard integer programming. In the next, We propose a low computational complexity method for blockage detection.
At the first step, we assume only one antenna is blocked, resulting in $N_{\text{S}}+1$ different candidate masking vectors ${\mv_{<0>}, \ldots, \mv_{<N_\text{S}>}}$ (where $\mv_{<i>}$ has all ones except the $i$-th element as zero, and $\mv_{<0>} = \mathbf{1}_{N}$). By calculating the cost based on~\eqref{eq_blockage_cost_function}, the masking vector with the lowest value will be considered, and the blockage index is denoted as $i_C$. Note that if the vector $m_{<0>}$ has the lowest cost, no blockage happens.
Starting from $\mv_{<i_C>}$, we evaluate the masking vectors by including left elements $\mv_{<i:i_C>}$ (where $\mv_{<i:j>}$ has all ones except the $i$-th to $j$-th entries), until the cost function is not decreasing or reaching the first element and we denote the leftmost blocked antenna as $i_L$.
Similar can be performed for $\mv_{<i_L:i>}$ by incorporating the right antennas, and $\mv_{<i_L:i_R>}$ will be taken as the masking vector. Once the masking vector has been obtained, the localization and sensing results can also be refined iteratively. The algorithm can be found in Algorithm~\ref{alg_partial_blockage_detection}.

% \begin{enumerate}
%     \item Assume only one antenna is blocked.
%     \item Start from point n that has the lowest cost $\Jm(\hat \sv, \mv)$.
%     \item Check cost by including left antennas in the masking vector $\mv$, e.g., $\mv_{[n-1, n]} = 0$
%     \item Stop when cost(nL-1, n) $>$ cost(nL, n)
%     \item Check cost by including right antennas, e.g., [nL, n+1]
%     \item Stop when cost(nL, nR+1) $>$ cost(nL, nR)
%     \item Go back to 2 until converge.
% \end{enumerate}

\begin{algorithm}[h]
\small
\caption{Partial Blockage Detection}
\label{alg_partial_blockage_detection}
\begin{algorithmic}[1]
\State \textbf{--- \textit{Heuristic Detection} ---}
\State Input: $\tilde \Ym$, $\hat \sv$
\State $i_\text{C} = \argmin_i J(\hat \sv, \mv_{<i>})$ ($i = 0, 1, \ldots, N$) using \eqref{eq_blockage_cost_function}
% , where $\mv_{<i>}$ has all ones except the $i$-th entry as zero, and $\mv_{<0>} = \mathbf{1}_{N}$
\If{$i_\text{C} \neq 0$}
    \State $i_\text{L} \leftarrow i_\text{C}$, $i_\text{R} \leftarrow i_\text{C}$
    \If{$i_\text{L} \neq 1$}
        \State {$i_\text{L} = \argmin_i J(\hat \sv, \mv_{<i:i_{\text{C}}>})$ ($i = 1, \ldots, i_{\text{C}}-1$) 
        }
        % , where $\mv_{<i:j>}$ has all ones except the $i$-$j$ entries}
    \EndIf
    \If{$i_\text{R} \neq N$}
        \State {$i_\text{R} = \argmin_i J(\hat \sv, \mv_{<i_{\text{L}}:i>})$ ($i=i_{\text{C}}+1, \ldots, N$)}
    \EndIf
\EndIf
% \For{$\ell = 1$ to $L$}
%     \State Estimate $\hat{\tau}_\ell$ using \eqref{eq:coarse_ris_H} and \eqref{eq_coarse_delay_RIS}.
%     \State Estimate $\hat \xi_\ell$ and $\hat \zeta_\ell$ using 2D grid search in \eqref{eq:coarse_ris_xizeta}.
%     \State $\hat{\etav}_{\text{N}, \ell}$ $\leftarrow$ $[\hat{\xi}_\ell, \hat{\zeta}_\ell, \hat{\tau}_\ell]^\top$.
% \EndFor
\Return $\hat{\mv} \leftarrow \mv_{<i_{\text{L}}:i_{\text{R}}>}$
\State \textbf{--- \textit{Iterative Refinement (Optional)} ---}
% \State Input: 
\State Estimate the state vector $\hat \sv$ by solving \eqref{eq_mle} with $\yv$, $\hat{\mv}$ as inputs.
\State Repeat steps 1-8 for a refined $\hat \mv$
% \Return refined $\hat{\etav}_\text{N} = [\hat{\eta}_{\text{N}, 0}, \hat{\etav}_{\text{N}, 1}, \ldots, \hat{\etav}_{\text{N}, L}]^\top$.
\end{algorithmic}
\end{algorithm}

\section{Simulation}

% \subsection{Simulation Setup}
We consider a $ \unit[28]{GHz}$ system with a bandwidth $W=\unit[200]{MHz}$, and $K=10$ subcarriers for localization. A single-antenna \ac{ue} located at $\pv_0 = [2, 4]^\top\unit[]{m}$ and one SP located at $\pv_1 = [2, -2]^\top\unit[]{m}$, and the array size is set as $N = 100$. For reference, the Fresnel distance and Fraunhofer distance are $\unit[2.2]{m}$ and $\unit[50]{m}$, respectively. We assume $M=4$ RFCs are connected to $S=4$ \acp{sa} (e.g., the first \ac{sa} contains antennas $1$-$25$), and the number of transmissions is set as $G=25$. The RCS coefficient for the SP is set as $c_\text{RCS} = \unit[0.5]{m^2}$, noise figure is set as $\unit[13]{dBm}$, and the noise \ac{psd} is set as $\unit[-173.855]{dBm/Hz}$.

\subsection{Evaluation of the Estimator }
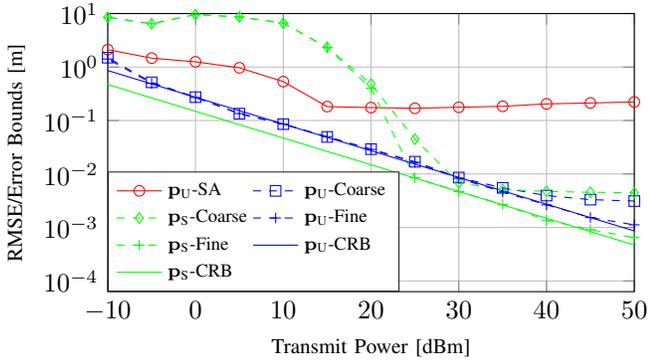
\begin{figure}[t!]
\begin{minipage}[h]{0.78\linewidth}
\centering
% \centerline{\includegraphics[width=0.99\linewidth]{Figures/fig_3.png}}
\begin{tikzpicture}

\begin{axis}[%
width=7cm,
height=3.7cm,
at={(0in,0in)},
scale only axis,
xmin=-10,
xmax=50,
xlabel style={font=\color{white!15!black},font=\footnotesize},
xlabel={Transmit Power [dBm]},
ymode=log,
ymin=6.3e-05,
ymax=10,
yminorticks=false,
ylabel style={font=\color{white!15!black},font=\footnotesize},
ylabel={$\text{RMSE/Error Bounds [m]}$},
axis background/.style={fill=white},
xmajorgrids,
ymajorgrids,
% yminorgrids,
legend style={at={(0, 0)}, anchor=south west, legend cell align=left, align=left, font=\scriptsize, legend columns=2, draw=white!15!black}
]
\addplot [color=red, mark size=2.0pt, mark=o, mark options={solid, red}]
  table[row sep=crcr]{%
-20	4.41319392104107\\
-15	3.0000876389964\\
-10	2.11080761512461\\
-5	1.46578941963627\\
0	1.25216551157063\\
5	0.966266753154699\\
10	0.534501700375304\\
15	0.182209228657012\\
20	0.17434923370892\\
25	0.169234988114999\\
30	0.176920051058823\\
35	0.184239725934509\\
40	0.204671238788871\\
45	0.212804985763281\\
50	0.222410548099171\\
};
\addlegendentry{$\pv_\text{U}$-SA}

\addplot [color=blue, dashed, mark size=2.0pt, mark=square, mark options={solid, blue}]
  table[row sep=crcr]{%
-20	4.40363943322781\\
-15	2.79616343887099\\
-10	1.49739750190313\\
-5	0.517714994651579\\
0	0.273183846225113\\
5	0.134320758685142\\
10	0.0856802047171063\\
15	0.0492641033705898\\
20	0.0291053566895087\\
25	0.0169609535875172\\
30	0.00857597313800324\\
35	0.00550280288309766\\
40	0.00394923445271583\\
45	0.00331345143475006\\
50	0.00311582554731855\\
};
\addlegendentry{$\pv_\text{U}$-Coarse}

\addplot [color=green, dashed, mark size=2.0pt, mark=diamond, mark options={solid, green}]
  table[row sep=crcr]{%
-15	4.62521969166677\\
-10	8.49399197196706\\
-5	6.42202166027007\\
0	9.56898119150902\\
5	8.6000386728662\\
10	6.62455707601758\\
15	2.33846634256473\\
20	0.48226057482434\\
25	0.044795008172576\\
30	0.00683546877480002\\
35	0.00514695918791166\\
40	0.00472431501708773\\
45	0.00454599557754727\\
50	0.00446003525163268\\
};
\addlegendentry{$\pv_\text{S}$-Coarse}

\addplot [color=blue, dashed, mark size=2.0pt, mark=+, mark options={solid, blue}]
  table[row sep=crcr]{%
-20	4.43044653015259\\
-15	2.80885045883595\\
-10	1.41096818223554\\
-5	0.494022413247659\\
0	0.269108279744256\\
5	0.134761936590862\\
10	0.0860346353748365\\
15	0.0498128272385674\\
20	0.0287598776689914\\
25	0.0165384709256327\\
30	0.00830808770131192\\
35	0.00459223952728711\\
40	0.00264852965431059\\
45	0.00153230216783813\\
50	0.00111852439280876\\
};
\addlegendentry{$\pv_\text{U}$-Fine}
% , line width=1.pt
\addplot [color=green, dashed, mark size=2.0pt, mark=+, mark options={solid, green}]
  table[row sep=crcr]{%
-15	4.64988891705464\\
-10	8.52171383244601\\
-5	6.44239558124915\\
0	9.5957278874457\\
5	8.64741058103881\\
10	6.64857709331158\\
15	2.31294435657225\\
20	0.400585615379671\\
25	0.00849475110770355\\
30	0.00468322760564079\\
35	0.00267879526327123\\
40	0.00136601622995724\\
45	0.000909579474787181\\
50	0.000647510886655765\\
};
\addlegendentry{$\pv_\text{S}$-Fine}

\addplot [color=blue]
  table[row sep=crcr]{%
-20	2.71953621767725\\
-15	1.52930760259594\\
-10	0.859992844717281\\
-5	0.483609533175642\\
0	0.271953618336603\\
5	0.152930761203781\\
10	0.0859992877117471\\
15	0.0483609506310146\\
20	0.0271953618212505\\
25	0.01529307578532\\
30	0.0085999285011394\\
35	0.00483609506803906\\
40	0.00271953609489461\\
45	0.00152930763011638\\
50	0.000859992832652176\\
};
\addlegendentry{$\pv_\text{U}$-CRB}

\addplot [color=green]
  table[row sep=crcr]{%
-20	1.48766148386663\\
-15	0.836573530330372\\
-10	0.470439867519103\\
-5	0.264547778641317\\
0	0.148766148364122\\
5	0.0836573530404746\\
10	0.0470439867757097\\
15	0.0264547778442471\\
20	0.0148766148376489\\
25	0.00836573530178417\\
30	0.00470439867578852\\
35	0.00264547778473681\\
40	0.0014876614831852\\
45	0.000836573530447073\\
50	0.000470439867454618\\
};
\addlegendentry{$\pv_\text{S}$-CRB}

% \addplot [color=black]
%   table[row sep=crcr]{%
% -20	0.134725369075725\\
% -15	0.0757616425810348\\
% -10	0.042603902485884\\
% -5	0.0239579349813454\\
% 0	0.0134725369075821\\
% 5	0.00757616425817497\\
% 10	0.00426039024865025\\
% 15	0.00239579349824537\\
% 20	0.00134725369065051\\
% 25	0.000757616425832042\\
% 30	0.000426039024866291\\
% 35	0.000239579349812321\\
% 40	0.000134725369058591\\
% 45	7.57616425771839e-05\\
% 50	4.26039024853691e-05\\
% };
% \addlegendentry{CRB-NF (Sync)}

\end{axis}

\end{tikzpicture}%
% (\textit{Figure elements by macrovector on Freepik})} 
\vspace{-8mm}
\end{minipage}
\caption{RMSE of the estimated UE and SP positions.}
\label{fig_2_simulation_estimators}
\end{figure}

We evaluate the \ac{rmse} of the estimated \ac{ue} and SP positions with different estimators, as shown in Fig.~\ref{fig_2_simulation_estimators}. The figure shows that the \ac{sa}-based solution in~\eqref{eq_coarse_PU_SA} provides limited accuracy. With coarse estimation based on~\eqref{eq_coarse_UE_position}, $\pv_\text{U}$ estimation can be improved but will deviate from the bound at high SNR due to the ignorance of weak NLOS path. Similar can be found in SP position results, where the deviation happens because of the residual error of LOS removal. However, with joint optimization of both paths based on~\eqref{eq_mle}, the RMSEs can attain the bound. Note that we assume the number of SPs is known, and a wrong number of $L$ will degrade localization and sensing performance. Compared with the RMSE of UE position that can provide good estimations at $\unit[-10]{dBm}$, satisfactory sensing results can only be obtained at high frequencies due to the weak signals that are easy to be interfered with by the LOS signal and background noise. Finally, the CRB of $\pv_\text{U}$ and $\pv_\text{S}$ are plotted in the black curve to benchmark the estimators.

\subsection{Effect of Partial Blockage}
To evaluate the effect of partial blockage on localization in terms of the biased term. The UE is located on a grid within a $\unit[5\times 16]{m^2}$ with an $\unit[1]{m}$ interval. The pseudotrue locations are plotted in red, with the segment representing the Biased term. In (a), where only one antenna is blocked, we find that the effect is ignorable. When more antennas are blocked, we can see that the target at a large angle of \ac{aoa} will be affected more severely, and the biased term at different locations shows a different pattern. When a partial blockage happens at different locations, we can see that the blockage of middle antennas has less effect on localization, indicating less bias introduced. As a result, detecting partial blockage and identifying blocked antennas can help in localization and sensing.

\begin{figure}[h]
\centering
% \hspace{-0.2cm} 
\begin{minipage}[b]{0.47\linewidth}
\centering
\centerline{\includegraphics[width=0.95\linewidth]{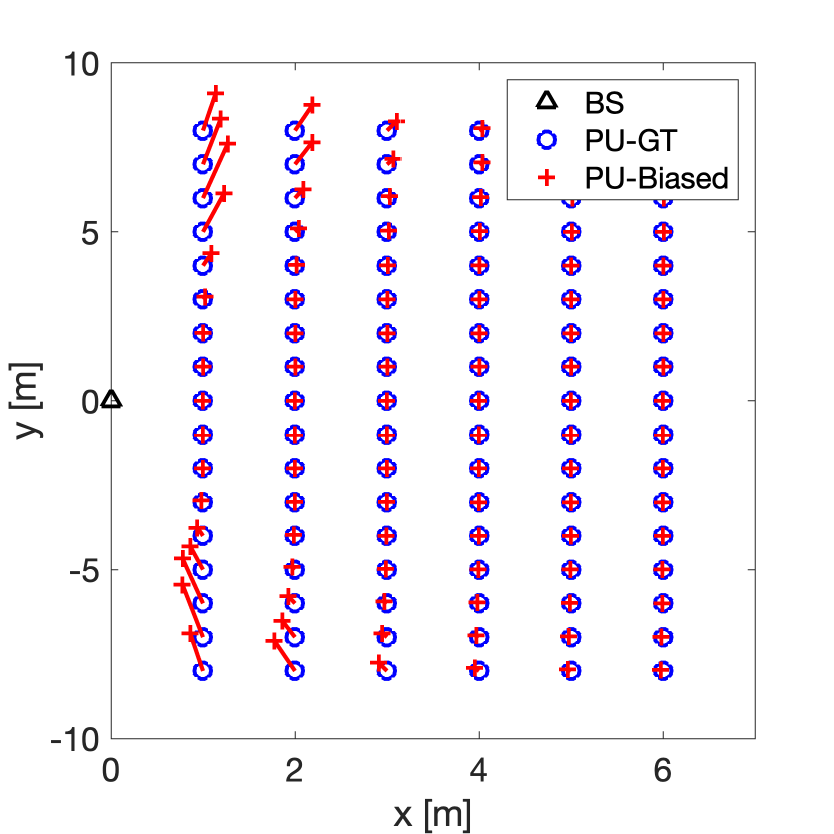}}
% \vspace{-0.5 cm}
\centerline{\footnotesize{(a)}} \medskip
\end{minipage}
% \vspace{0.5cm}
\;
\begin{minipage}[b]{0.47\linewidth}
\centering
\centerline{\includegraphics[width=0.95\linewidth]{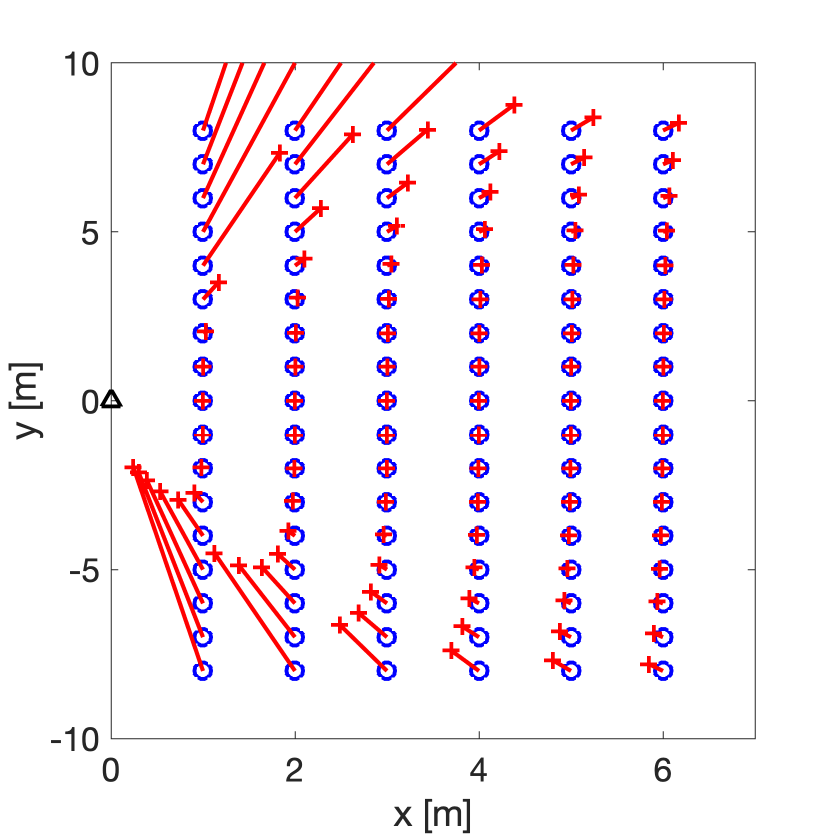}}
% \include{Figures_Tikz/2}
% \vspace{-0.5 cm}
\centerline{\footnotesize{(b)}} \medskip
\end{minipage}
\begin{minipage}[b]{0.47\linewidth}
\centering
\centerline{\includegraphics[width=0.95\linewidth]{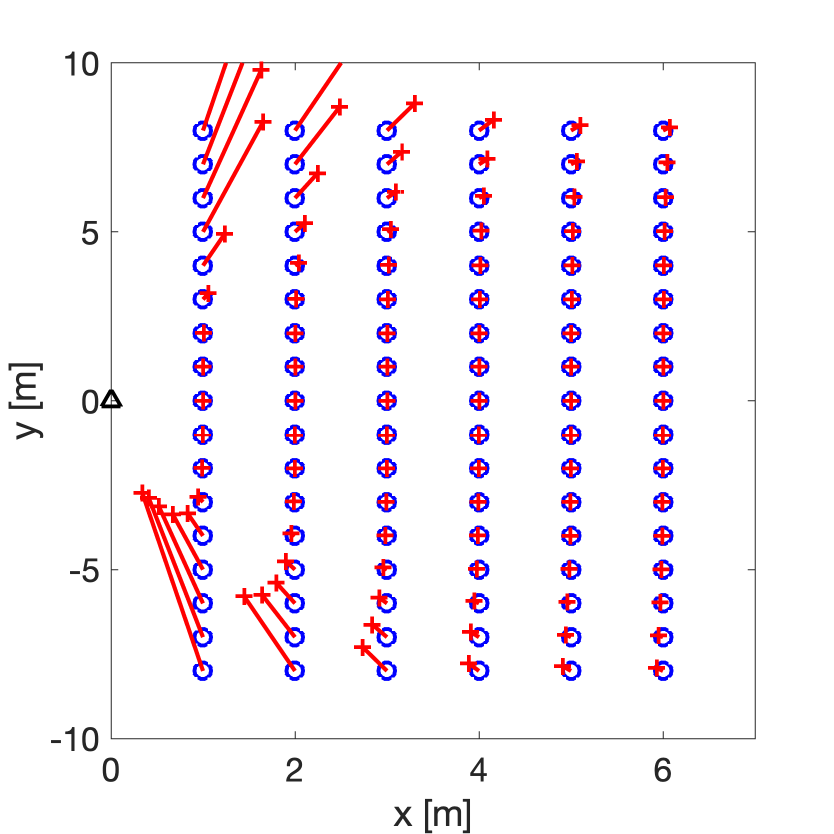}}
% \vspace{-0.5 cm}
\centerline{\footnotesize{(c)}} \medskip
\end{minipage}
% \vspace{0.5cm}
\;
\begin{minipage}[b]{0.47\linewidth}
\centering
\centerline{\includegraphics[width=0.95\linewidth]{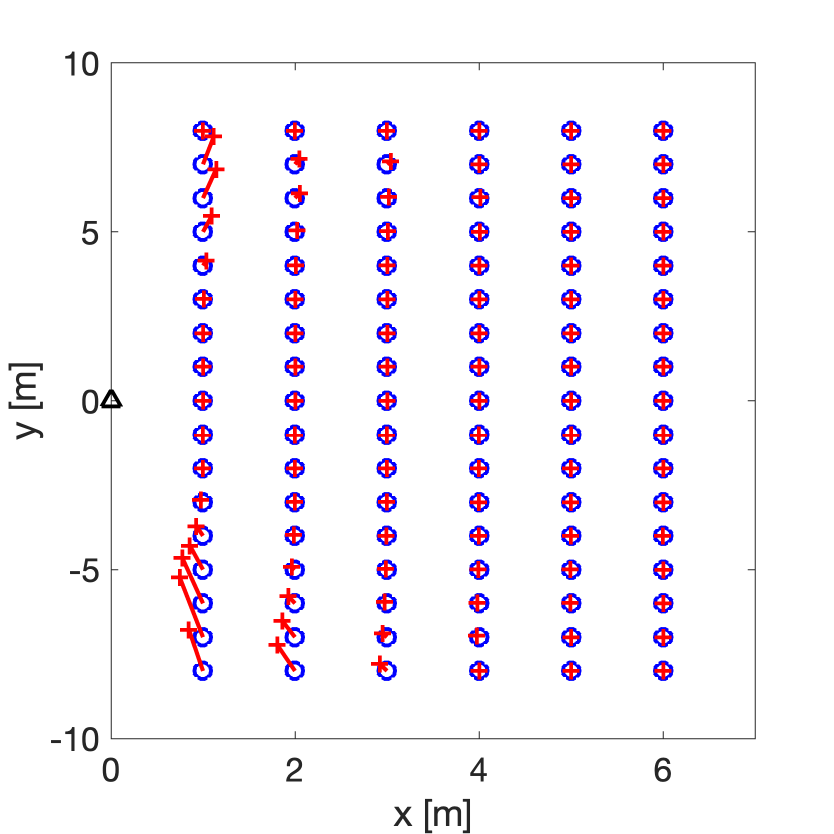}}
% \include{Figures_Tikz/2}
    % \vspace{-0.5 cm}
\centerline{\footnotesize{(d)}} \medskip
\end{minipage}
\caption{The effect of partial blockage on localization under different scenarios
(blue circle indicates the ground truth and the red cross indicates the biased
estimation): (a) antenna $100$ is blocked; (b) antennas $96$-$100$ are blocked;
(c) antennas $76$-$80$ are blocked; (d) antennas $56$-$60$ are blocked.}
\label{fig_3_partial_blockage_detection}
% \vspace{0.5cm}
\end{figure}
% \vspace{-0.5cm}

\subsection{Evaluation of Partial Blockage Detection Algorithm}
\begin{figure}[t!]
\begin{minipage}[h]{0.78\linewidth}
\centering
% \centerline{\includegraphics[width=0.99\linewidth]{Figures/fig_3.png}}
\begin{tikzpicture}

\begin{axis}[%
width=7cm,
height=3.8cm,
at={(0in,0in)},
scale only axis,
xmin= 0,
xmax= 25,
xlabel style={font=\color{white!15!black},font=\footnotesize},
xlabel={Antenna Index},
% ymode=log,
ymin=0,
ymax=0.25,
yminorticks=false,
ylabel style={font=\color{white!15!black},font=\footnotesize},
ylabel={$\text{Cost Value}$},
axis background/.style={fill=white},
xmajorgrids,
ymajorgrids,
% yminorgrids,
legend style={at={(0, 1)}, anchor=north west, legend cell align=left, align=left, font=\scriptsize, legend columns=1, draw=white!15!black}
]

\addplot [color=blue, mark=o, mark options={solid, blue}]
  table[row sep=crcr]{%
1	0.0181099350191147\\
2	0.0162201580579534\\
3	0.0149536071294783\\
4	0.013868022847786\\
5	0.0119310589407083\\
6	0.00971988967989115\\
7	0.0109240316544428\\
8	0.0110249480202971\\
9	0.0107276388870353\\
10	0.00869023307382199\\
11	0.0109153905904399\\
12	0.0129863140915845\\
13	0.0142886856605843\\
14	0.0158067225081783\\
15	0.0172268518729452\\
16	0.0188587617201897\\
17	0.0196928766431127\\
18	0.0208430714929188\\
19	0.0219704598920394\\
20	0.0230268984493799\\
21	0.0240346211624862\\
22	0.0251043579478562\\
23	0.0264235758249719\\
24	0.0269026862734809\\
25	0.0275675376736431\\
};
\addlegendentry{$G=5, P=\unit[20]{dBm}$ (6-10)}

\addplot [color=red, mark=square, mark options={solid, red}]
  table[row sep=crcr]{%
1	0.0396563202860024\\
2	0.0366186074578695\\
3	0.0333442865460972\\
4	0.0291005623319724\\
5	0.0241786345062798\\
6	0.0177485308415167\\
7	0.0219845247553846\\
8	0.023320397842305\\
9	0.0228743121579559\\
10	0.0198066693016881\\
11	0.0258723918338482\\
12	0.0303096277012271\\
13	0.0347474070174656\\
14	0.0385134571800332\\
15	0.0417977541161563\\
16	0.0448103002693968\\
17	0.0472102423430996\\
18	0.0495783249952322\\
19	0.0519369173867193\\
20	0.0541659342100788\\
21	0.0565479402855331\\
22	0.0586075304500809\\
23	0.0607831693162465\\
24	0.0628644294171387\\
25	0.0649553390129757\\
};
\addlegendentry{$G=20, P=\unit[20]{dBm}$ (6-10)}

\addplot [color=green, mark=diamond, mark options={solid, green}]
  table[row sep=crcr]{%
1	0.124229872902903\\
2	0.114711450411292\\
3	0.104251789249376\\
4	0.0907948398976552\\
5	0.074909783794555\\
6	0.0527416109418286\\
7	0.0680266927769998\\
8	0.0720738737177306\\
9	0.0701587170929263\\
10	0.0582895478705951\\
11	0.079643108762922\\
12	0.0942505521993156\\
13	0.108519960575355\\
14	0.120610509733422\\
15	0.131195656640609\\
16	0.140812913041986\\
17	0.148590566791765\\
18	0.156102109770984\\
19	0.163730238041128\\
20	0.170842954288094\\
21	0.178259560085201\\
22	0.184606183338888\\
23	0.191573350585471\\
24	0.198329671483224\\
25	0.205048117115157\\
};
\addlegendentry{$G=20, P=\unit[30]{dBm}$ (6-10)}

\addplot [color=black, mark=triangle, mark options={solid, black}]
  table[row sep=crcr]{%
1	0.1635389344232\\
2	0.157080372150201\\
3	0.148836564611543\\
4	0.140858910742075\\
5	0.132512754539115\\
6	0.122461844868434\\
7	0.113703995218364\\
8	0.103773145662384\\
9	0.0914323820919425\\
10	0.076760010626467\\
11	0.05619004986471\\
12	0.07212994063575\\
13	0.0767839622229291\\
14	0.0748613201191801\\
15	0.0629458309184811\\
16	0.082498419935931\\
17	0.0951768643622761\\
18	0.108123672549379\\
19	0.11754778152275\\
20	0.128247081580531\\
21	0.138358557679656\\
22	0.147026237159229\\
23	0.154489641968033\\
24	0.161983412915174\\
25	0.169173591947463\\
};
\addlegendentry{$G=20, P=\unit[30]{dBm}$ (11-15)}

\end{axis}

\end{tikzpicture}%
% (\textit{Figure elements by macrovector on Freepik})}
\vspace{-5mm}
\end{minipage}
\vspace{-0.5cm}
\caption{Cost function of the heuristic algorithm (with partial blockage in antennas $6$-$10$ and $11$-$15$).}
\label{fig_4_cost_function_heuristic_algorithm}
\end{figure}
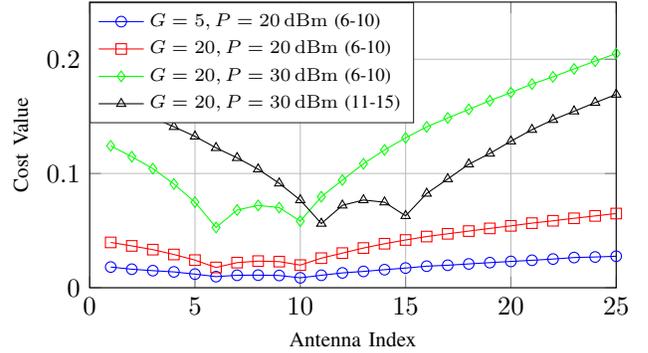

The cost function after performing steps $1$-$9$ in Algorithm~\ref{alg_partial_blockage_detection} is shown in Fig.~\ref{fig_4_cost_function_heuristic_algorithm} (averaged with $100$ times Monte Carlo simulations). From the benchmark scenario with a limited number of transmission and transmit power (blue curve with circle markers), the averaged cost values are close to each other, indicating an inaccurate detection is likely to happen. With increased transmissions (red curves with square markers), two local minima (antenna index $6$ and $10$) are more visible, indicating the boundary of the blocked antennas. The averaged cost functions with higher transmit power for different blocked antennas (green curve with diamond markers and black curve with triangle markers) are also shown in the figure, indicating these setups are likely to be successfully detected.

% \vspace{-4mm}

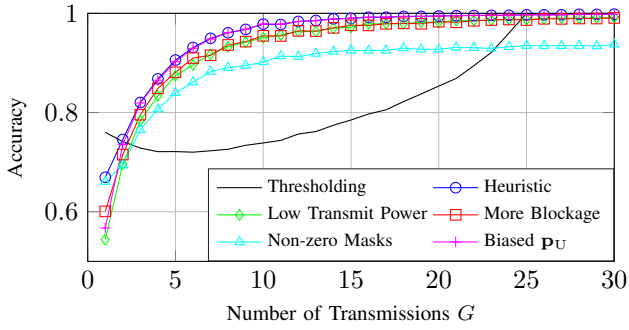
\begin{figure}[t!]
\begin{minipage}[h]{0.78\linewidth}
\centering
% \centerline{\includegraphics[width=0.99\linewidth]{Figures/fig_3.png}}
\definecolor{mycolor1}{rgb}{0.00000,1.00000,1.00000}%
\definecolor{mycolor2}{rgb}{1.00000,0.00000,1.00000}%
\begin{tikzpicture}

\begin{axis}[%
width=7cm,
height=3.3cm,
at={(0in,0in)},
scale only axis,
xmin= 0,
xmax= 30,
xlabel style={font=\color{white!15!black},font=\footnotesize},
xlabel={Number of Transmissions $G$},
% ymode=log,
ymin=0.5,
ymax=1,
yminorticks=false,
ylabel style={font=\color{white!15!black},font=\footnotesize},
ylabel={$\text{Accuracy}$},
axis background/.style={fill=white},
xmajorgrids,
ymajorgrids,
% yminorgrids,
legend style={at={(1, 0)}, anchor=south east, legend cell align=left, align=left, font=\scriptsize, legend columns=2, draw=white!15!black}
]

\addplot [color=black]
  table[row sep=crcr]{%
1	0.759999999999999\\
2	0.741479999999999\\
3	0.72828\\
4	0.72112\\
5	0.72128\\
6	0.72008\\
7	0.72292\\
8	0.726\\
9	0.733960000000001\\
10	0.73872\\
11	0.74384\\
12	0.75668\\
13	0.761640000000001\\
14	0.7744\\
15	0.784920000000001\\
16	0.797040000000001\\
17	0.805440000000001\\
18	0.822160000000001\\
19	0.83748\\
20	0.85332\\
21	0.86888\\
22	0.894439999999999\\
23	0.921119999999999\\
24	0.9602\\
25	1\\
26	1\\
27	1\\
28	1\\
29	1\\
30	1\\
};
\addlegendentry{Thresholding}

\addplot [color=blue, mark=o, mark options={solid, blue}]
  table[row sep=crcr]{%
1	0.66952\\
2	0.74572\\
3	0.82036\\
4	0.86784\\
5	0.9058\\
6	0.93108\\
7	0.9496\\
8	0.96076\\
9	0.96756\\
10	0.97816\\
11	0.9774\\
12	0.98364\\
13	0.98556\\
14	0.9888\\
15	0.98972\\
16	0.99208\\
17	0.99212\\
18	0.99368\\
19	0.99376\\
20	0.9946\\
21	0.99516\\
22	0.99544\\
23	0.99592\\
24	0.99604\\
25	0.9972\\
26	0.99688\\
27	0.99712\\
28	0.998\\
29	0.99788\\
30	0.99824\\
};
\addlegendentry{Heuristic}

\addplot [color=green, mark=diamond, mark options={solid, green}]
  table[row sep=crcr]{%
1	0.54384\\
2	0.696\\
3	0.78396\\
4	0.83364\\
5	0.8744\\
6	0.89584\\
7	0.91744\\
8	0.93248\\
9	0.9416\\
10	0.95096\\
11	0.95612\\
12	0.96332\\
13	0.96348\\
14	0.97012\\
15	0.97188\\
16	0.97892\\
17	0.97908\\
18	0.97908\\
19	0.9822\\
20	0.9822\\
21	0.98528\\
22	0.98548\\
23	0.98624\\
24	0.98808\\
25	0.9874\\
26	0.9894\\
27	0.99044\\
28	0.99028\\
29	0.99072\\
30	0.99184\\
};
\addlegendentry{Low Transmit Power}

% \addplot [color=red, mark=square, mark options={solid, red}]
%   table[row sep=crcr]{%
% 1	0.59512\\
% 2	0.692559999999999\\
% 3	0.789479999999999\\
% 4	0.848239999999999\\
% 5	0.894599999999999\\
% 6	0.925959999999999\\
% 7	0.931799999999999\\
% 8	0.94812\\
% 9	0.96132\\
% 10	0.96656\\
% 11	0.97252\\
% 12	0.97516\\
% 13	0.98084\\
% 14	0.98164\\
% 15	0.98296\\
% 16	0.98712\\
% 17	0.987\\
% 18	0.98808\\
% 19	0.98928\\
% 20	0.9906\\
% 21	0.99148\\
% 22	0.99276\\
% 23	0.9932\\
% 24	0.9938\\
% 25	0.99372\\
% 26	0.9944\\
% 27	0.99448\\
% 28	0.99536\\
% 29	0.99568\\
% 30	0.99636\\
% };
% \addlegendentry{High Reflection}

\addplot [color=red, mark=square, mark options={solid, red}]
  table[row sep=crcr]{%
1	0.60076\\
2	0.71524\\
3	0.79568\\
4	0.84888\\
5	0.88104\\
6	0.90984\\
7	0.91536\\
8	0.93624\\
9	0.94296\\
10	0.95464\\
11	0.95436\\
12	0.96476\\
13	0.96408\\
14	0.97072\\
15	0.97596\\
16	0.97564\\
17	0.9788\\
18	0.97984\\
19	0.97896\\
20	0.98304\\
21	0.98192\\
22	0.985\\
23	0.98652\\
24	0.98764\\
25	0.98648\\
26	0.989\\
27	0.9888\\
28	0.98956\\
29	0.98952\\
30	0.99072\\
};
\addlegendentry{More Blockage}

\addplot [color=mycolor1, mark=triangle, mark options={solid, mycolor1}]
  table[row sep=crcr]{%
1	0.66064\\
2	0.69364\\
3	0.76556\\
4	0.80684\\
5	0.83972\\
6	0.8612\\
7	0.88268\\
8	0.89044\\
9	0.89484\\
10	0.90192\\
11	0.9138\\
12	0.91264\\
13	0.91896\\
14	0.923\\
15	0.92532\\
16	0.92504\\
17	0.92536\\
18	0.92984\\
19	0.9274\\
20	0.92756\\
21	0.93112\\
22	0.93084\\
23	0.92936\\
24	0.93264\\
25	0.93484\\
26	0.93468\\
27	0.93448\\
28	0.93472\\
29	0.93416\\
30	0.93756\\
};
\addlegendentry{Non-zero Masks}

\addplot [color=mycolor2, mark=+, mark options={solid, mycolor2}]
  table[row sep=crcr]{%
1	0.56728\\
2	0.73528\\
3	0.81928\\
4	0.86512\\
5	0.90252\\
6	0.92944\\
7	0.94832\\
8	0.96104\\
9	0.96796\\
10	0.97856\\
11	0.97808\\
12	0.98356\\
13	0.9852\\
14	0.98816\\
15	0.99024\\
16	0.99228\\
17	0.99236\\
18	0.99384\\
19	0.99388\\
20	0.99456\\
21	0.99552\\
22	0.99588\\
23	0.99604\\
24	0.99612\\
25	0.99716\\
26	0.99668\\
27	0.99704\\
28	0.99816\\
29	0.9978\\
30	0.99836\\
};
\addlegendentry{Biased $\pv_\mathrm{U}$}

\end{axis}

\end{tikzpicture}%
\vspace{-5mm}
\end{minipage}
\vspace{-0.5cm}
\caption{The evaluation of the \ac{pbd} methods under different scenarios.}
\label{fig_3_partial_blockage_detection}
\end{figure}
% \vspace{-10mm}

We take the first \ac{sa} with antenna index $1$-$25$) to evaluate two different partial blockage methods with different known UE locations (i.e., a true location and a biased location). The default parameters are set as $P = \unit[20]{dBm}$, $c = \unit[0.5]{m^2}$, with antennas $5$-$10$ in the \ac{sa} blocked as the benchmark. Other scenarios such as lower power ($P = \unit[0]{dBm}$), {non-zero masks (the coefficient follows $\mathcal{CN}(0.2, 0.2)$ instead of $0$)} and more blockage (antennas $5$-$15$ are blocked), biased $\pv = [2.0691, 4.1377]^\top \unit[]{m}$ (based on~\eqref{eq_pseudotrue_parameter}) are also evaluated. We can see that the thresholding method (black curve) only works for a sufficient number of transmissions ($G \ge 25$), and accuracy decreases with a reduced value of $G$. When $G=1$, no blocked antennas can be detected, resulting in a constant value of $1-6/25 = 0.76$. However, the heuristic method (blue curve) shows a much-improved performance. The results also show that a lower transmit power, more blockages, and a wrong assumption of the masking vector decrease detection accuracy. However, a biased position estimate has a limited effect on the accuracy due to the fact that the biased position minimizes the \ac{kld}, providing similar signal observations as the ground truth positions.
\section{Conclusion}
In this work, we studied a SIMO uplink joint synchronization, localization, and sensing problem in the \ac{nf} scenario with a single \ac{bs}, which was previously impossible with a far-field model. A subarray-based coarse localization and sensing algorithm is performed, followed by an MLE-based refinement. We also analyzed the impact of localization under partial blockage and proposed a heuristic blockage detection algorithm to mitigate the effect of blockage. However, the reflection and diffraction of the blockage itself are not considered, and joint localization and \ac{pbd} in 3D space need to be considered in future work.

\section*{Acknowledgment}
This work has been supported by the SNS JU project 6G-DISAC under the EU’s Horizon Europe research and innovation programme under Grant Agreement No 101139130.
% This work was supported, in part, by the European Commission through the H2020 project Hexa-X (Grant Agreement no. 101015956) and by the MSCA-IF grant 888913 (OTFS-RADCOM), and by Academy of Finland Profi-5 (n:o 326346) and ULTRA (n:o 328215) projects.

\vspace{-1mm}
\bibliographystyle{IEEEtran}
\bibliography{ref}

% \newpage

 

% \end{IEEEbiography}



\vfill

\end{document}